\documentclass[final,5p,times,twocolumn]{elsarticle}

\usepackage{graphicx}
\usepackage{dcolumn}
\usepackage{bm}
\usepackage{longtable}
\usepackage{lscape}
\usepackage{multirow}
\usepackage{booktabs}
\usepackage{amsmath}
\usepackage{float}
\usepackage{rotating}
\usepackage{caption}
\usepackage{placeins}
\usepackage{multicol}
\usepackage{xcolor}
\usepackage[colorlinks=true,linkcolor=blue,citecolor=blue,urlcolor=blue]{hyperref}
\usepackage{miller}

\journal{Acta Materialia}

\begin{document}

\begin{frontmatter}

\title{Solute Co-Segregation Mechanisms at Low-Angle Grain Boundaries in Magnesium: \\A Combined Atomic-Scale Experimental and Modeling Study}

\author[1]{Risheng Pei\corref{cor1}}
\cortext[cor1]{Corresponding author}
\ead{pei@imm.rwth-aachen.de}
\affiliation[1]{organization={Institut für Metallkunde und Materialphysik, RWTH Aachen University},
    city={52056 Aachen},
    country={Germany}}

\author[5]{Joé Petrazoller}
\affiliation[5]{organization={Université de Lorraine, CNRS, Arts et Métiers, LEM3},
    city={57070 Metz},
    country={France}}

\author[2,6,fn1]{Achraf Atila}
\affiliation[2]{organization={Department of Materials Science and Engineering, Saarland University},
    city={66123 Saarbrücken},
    country={Germany}}
\affiliation[6]{organization={Federal Institute of Materials Research and Testing (BAM)},
city={Unter den Eichen 87, 12205, Berlin},
    country={Germany}}
\fntext[fn1]{Now at: Federal Institute of Materials Research and Testing (BAM), Unter den Eichen 87, 12205, Berlin, Germany}
    
\author[1]{Simon Arnoldi}
\author[3]{Lei Xiao}
\affiliation[3]{organization={College of Materials Science and Engineering, Fuzhou University},
    city={350108 Fuzhou},
    country={China}}
\author[4]{Xiaoqing Liu}
\affiliation[4]{organization={School of Physics and Electronics, Gannan Normal University},
    city={341000 Ganzhou},
    country={China}}
\author[1]{Hexin Wang}
\author[1]{Sandra Korte-Kerzel}
\author[5]{Stéphane Berbenni}
\author[5]{Thiebaud Richeton}
\author[5]{Julien Guénolé}
\author[1]{Zhuocheng Xie\corref{cor1}}
\ead{xie@imm.rwth-aachen.de}
\author[1]{Talal Al-Samman}

\begin{abstract}
Solute segregation at low-angle grain boundaries (LAGBs) critically affects the microstructure and mechanical properties of magnesium (Mg) alloys.  In modern alloys containing multiple substitutional elements, understanding solute-solute interactions at microstructural defects becomes essential for alloy design. This study investigates the co-segregation mechanisms of calcium (Ca), zinc (Zn), and aluminum (Al) at a LAGB in a dilute AZX010 Mg alloy by combining atomic-scale experimental and modeling techniques. Three-dimensional atom probe tomography (3D-APT) revealed significant segregation of Ca, Zn, and Al at the LAGB, with Ca forming linear segregation patterns along dislocation arrays characteristic of the LAGB. Clustering analysis showed increased Ca–Ca pairs at the boundary, indicating synergistic solute interactions. Atomistic simulations and elastic dipole calculations demonstrated that larger Ca atoms prefer tensile regions around dislocations, while smaller Zn and Al atoms favor compressive areas. These simulations also found that Ca–Ca co-segregation near dislocation cores is energetically more favorable than other solute pairings, explaining the enhanced Ca clustering observed experimentally. Thermodynamic modeling incorporating calculated segregation energies and solute-solute interactions accurately predicted solute concentrations at the LAGB, aligning with experimental data. The findings emphasize the importance of solute interactions at dislocation cores in Mg alloys, offering insights for improving mechanical performance through targeted alloying and grain boundary engineering.
\end{abstract}
\begin{keyword}
Atom probe tomography, atomistic simulation, grain boundary, co-segregation, magnesium alloy 
\end{keyword}

\end{frontmatter}

\section{Introduction}

Modern metallic alloys are typically  composed of multiple alloying elements that enhance desired material properties such as strength, toughness, ductility, and corrosion resistance. While these alloying elements play a crucial role in improving performance, they also introduce complex microstructural interactions across various length scales. Understanding these interactions, particularly at interfaces, is essential for the design and optimization of advanced structural materials with greater strength and ductility \cite{raabe2014grain}. Grain boundaries (GBs), as regions of high lattice disruption, act as thermodynamically favorable sinks for solute atoms due to their higher Gibbs free energy compared to the bulk \cite{lejček2008thermodynamics}. Consequently, the concentration of solutes at GBs can significantly exceed their solubility within the grain interior, sometimes by several orders of magnitude \cite{hofmann1996solute,basu2016role}. Lowering the energy state of the GB network through solute segregation reduces the driving force for capillary-driven grain growth during processing \cite{weissmuller1993alloy}. 

Beyond thermodynamic stabilization, segregated solute atoms within the GB region exhibit a kinetic effect on the movement of grain boundaries, known as solute drag \cite{lucke1957quantitative,cahn1962impurity}. This effect, influenced by solute diffusivity, retards the mobility of migrating boundaries \cite{huang2012effect} and can be strategically utilized to control the grain size distribution and texture \cite{hirsch2013superior,hadorn2012role}. In this context, recent studies have highlighted that GB solute segregation is influenced by the local grain boundary structure, resulting in a non-uniform distribution of solutes across different boundaries \cite{pei2023atomistic}. This variability leads to differing GB mobilities, which in turn facilitates the selective growth of specific texture components \cite{mouhib2022synergistic,zhang2023anisotropic}. In certain cases, solute segregation significantly alters the structural and chemical state of GBs, facilitating the formation of GB secondary phases \cite{raabe2013segregation} that further stabilize a fine grain size and enhance material properties through similar kinetic mechanisms like Zener pinning \cite{hillert1988inhibition}. 

While most segregation studies have traditionally focused on binary systems, recent research has shifted towards modern multi-component materials \cite{pei2021grain,zhao2019direct,zhang2022significantly,li2019segregation,li2020grain}. The presence of multiple alloying elements can significantly alter the GB segregation behavior, rendering the atomistic mechanisms at interfaces a complex and challenging topic to investigate \cite{xing2018solute}. Although solute segregation has been well established to enhance GB cohesion and bonding strength \cite{wu1994first,seah1975interface} or induce structural and chemical transitions, resulting in GB-stabilized phases known as complexions \cite{dillon2007complexion}, the study of interactions involving multiple solute species at interfaces remains relatively unexplored. For lightweight materials such as magnesium (Mg) alloys, dilute alloying with specific substitutional solute elements of larger and smaller atomic radii than Mg has been found to extend their utility by improving ductility and yield strength while mitigating mechanical anisotropy \cite{jung2015role, trang2018designing, pei2019superior}. In previous research on dilute systems like Mg-Gd-Zn, Mg-Ca-Zn, Mg-Al-Zn-Ca and Mg-Al-Zn-Ca-Y, it was observed that co-added solutes, particularly Zn, can influence and/or compete with the segregation tendencies of the other solute species to GB sites \cite{pei2019superior,pei2022synergistic, nie2008solute, jang2022comparative}. This was evident for elements like Gd and Ca with large atomic sizes. The effect of co-added solutes on segregation, known as synergistic co-segregation, has been recognized in these studies as a beneficial feature for generating unique textures that respond favorably to applied plastic deformation \cite{jung2015role, hirsch2013superior, mouhib2022synergistic}. Deeper insights into the synergistic nature of co-segregation have revealed that solute clustering and redistribution at GBs depend not only on minimizing elastic strain energy but also on the chemical binding strength with vacancies or co-existing solutes \cite{mouhib2024exploring, yi2023interplay}.

To gain comprehensive insights into the structure and chemistry of GBs with segregated solutes, it is essential to obtain precise and statistically significant experimental data at the atomic scale. Recent advances in characterization techniques, such as aberration-corrected scanning transmission electron microscopy (STEM) \cite{wang2011atom}, transmission Kikuchi diffraction (TKD) \cite{sneddon2016transmission}, and atom probe tomography (APT) \cite{gault2021atom}, have enabled detailed studies of atomic-scale chemical segregation and the crystallography associated with GBs. Utilizing atomic-resolution energy-dispersive X-ray spectroscopy (EDS) at low electron voltages to minimize beam damage, Zhao et al. \cite{zhao2019direct} uncovered a co-segregation pattern characterized by an alternating distribution of large Nd and small Ag atoms fully occupying the \hkl{10-11} and \hkl{10-12} twin boundaries in a Mg-Nd-Ag alloy. First-principles calculations indicated that this segregation pattern strengthens the twin boundary by enhancing bonding, thereby affecting its migration mechanism. 

In a study on the effect of co-segregation during annealing of a Mg-Al-Zn-Ca-Y alloy, Pei et al. \cite{pei2022synergistic} emphasized that the type and strength of segregation are crucial for restricting the preferential growth of recrystallized grains with a basal texture. Thermodynamically, the synergistic co-segregation of small Al and Zn atoms with large Ca atoms is more effective than mono-segregation in restricting grain growth because co-segregation results in a greater reduction in GB energy, rendering the boundary less mobile.

While high-resolution experimental techniques provide remarkable insights into interface segregation, numerous aspects remain elusive and challenging to capture experimentally. To bridge these gaps, complementary atomic-scale modeling, such as atomistic simulations and density functional theory (DFT) calculations, can be employed to gain a deeper understanding of the underlying processes and mechanisms beyond experimental limitations. A recent review of computational studies on solute segregation presented by Hu et al. \cite{hu2024computational} highlights significant efforts to investigate substitutional GB segregation. These efforts involve calculating the segregation energy of GB sites across a broad compositional space, revealing a strong correlation between segregation energy and site volume \cite{wang2024defects,huber2014atomistic,pei2019first}.

Although numerous studies have explored GB solute segregation in dilute binary solid solutions \cite{wagih2020learning,karkina2016solute,zhang2014twin, huber2017ab}, related research has also extended to concentrated solid solutions beyond the dilute limit, accounting for solute-solute \cite{wagih2020grain, zhang2024grain} or even multiple-solute interactions \cite{xing2018solute,cui2023effect,ganguly2024grain}. Additionally, incorporating vibrational entropy contributions has been shown to enhance predictions of GB segregation energy and solute concentration at elevated temperatures \cite{menon2024atomistic}. In the context of Mg alloys, most atomic-scale simulations using ab-initio or semi-empirical approaches have traditionally focused on individual GBs, particularly those with highly symmetric tilt configurations \cite{wang2024defects,huber2014atomistic,zhang2014twin, he2021unusual}. However, to truly understand co-segregation behavior in real, complex materials, it is necessary to consider a much larger number of GB sites and their combinations. This approach demands significantly more computational resources, especially when aiming for DFT-level accuracy \cite{menon2024atomistic}. In this regard, Wagih et al. demonstrated that highly symmetric tilt boundaries and low coincidence site lattice (CSL) GBs do not accurately represent polycrystalline GB environments, resulting in incorrect predictions of solute segregation behavior \cite{wagih2023can}. Building on this understanding, Pei et al. employed APT-informed atomistic simulations to investigate the critical phenomenon of inhomogeneous segregation behavior of Nd at general GBs in as-extruded Mg-Mn-Nd alloys \cite{pei2023atomistic}. Their findings revealed that this inhomogeneous segregation within the GB plane arises from local atomic arrangements rather than macroscopic crystallographic characteristics. 

The GB segregation energy spectrum and its local atomic environment have been extensively studied in randomly-textured face-centered cubic (FCC) polycrystals. However, polycrystalline Mg, which often exhibits a strong basal texture in conventional sheet alloys, has not received the same level of attention until recently. Recent research by the authors \cite{ganguly2024grain} has addressed this gap by calculating the segregation energy spectrum for basal-textured Mg polycrystals, emphasizing contributions from specific GB sites such as triple junctions. The study revealed a distinct bi-modal distribution that significantly differs from the conventional skew-normal distribution observed in randomly oriented polycrystals. 

In this study, solute co-segregation mechanisms at a LAGB in magnesium are investigated. The significance of examining solute segregation at LAGBs lies in their prevalence as subgrain structures in deformed materials and their distinct structural characteristics, which differ from those of random high-angle grain boundaries (HAGBs) and highly coherent grain boundaries that have been more extensively studied. LAGBs are characterized by discrete arrays of dislocations resulting from their small misorientation angles (ranging from 5° to 15°), creating unique sites for multiple solute atoms to co-segregate. This segregation notably impacts local strain fields and dislocation mobility, thereby influencing mechanical properties such as yield strength and ductility. By understanding how various solute species interact with different types of grain boundaries, namely LAGBs exhibiting solute-decorated dislocation cores, CSL GBs with precise atomic alignment, or HAGBs with more disordered structures, this research contributes to the development of a comprehensive understanding of how GB structure and characteristics affect solute behavior. 

\section{Methods}

\begin{figure*}[htbp!]
\centering
\includegraphics[width=\textwidth]{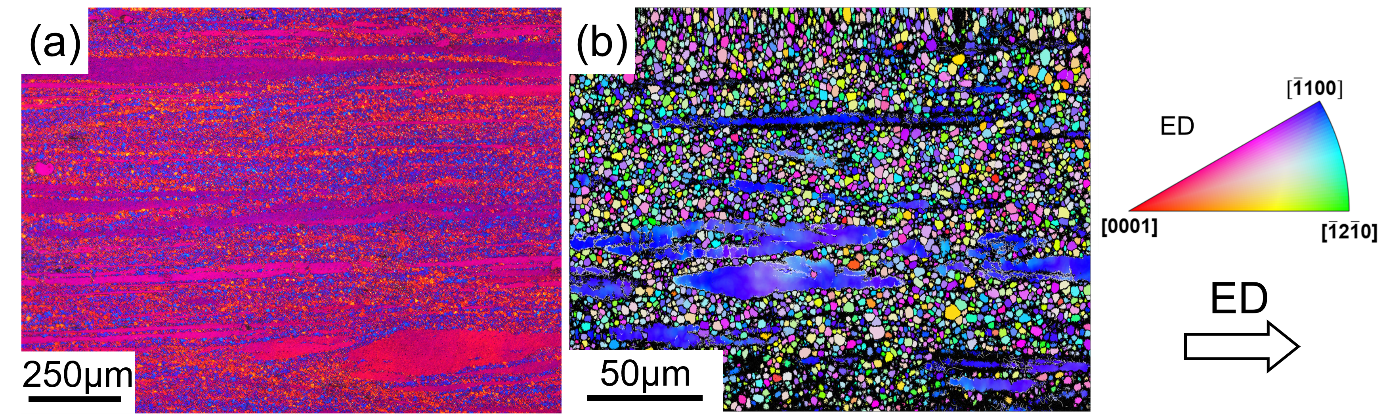}
\caption{Microstructure of as-extruded AZX010 alloy: (a) optical microscopic image; (b) EBSD map with IPF coloring with respect to the extrusion direction (ED).}
\label{fig1}
\end{figure*}

\subsection{Experiments}
The material investigated in this study was a  rare-earth free, lean Mg-0.23Al-1.00Zn-0.38Ca (wt.\%) alloy, designated as AZX010, in its as-extruded condition. Following gravity casting and a homogenization treatment at 420 °C for 24 hours, the alloy composition was characterized using inductively coupled plasma-optical emission spectrometry (ICP-OES). The alloy was then machined into extrusion billets with an initial diameter of  $D_0 = 45~\text{mm}$. Hot extrusion was carried out at 250 °C with a speed of 1.0~mm/s, producing extruded bars with a final diameter of $D_f = 9.0~\text{mm}$. This corresponded to an extrusion ratio of $\sim 1:25$ and a logarithmic strain of 3.22, calculated by $\ln\left(\frac{D_0}{D_f}\right)^2$.

The microstructure of the as-extruded sample, characterized using optical microscopy and orientation microscopy via electron backscatter diffraction (EBSD), consists of large elongated grains oriented along the extrusion direction (ED) and small recrystallized grains with an average grain size of 3.2 \(\mu\text{m}\), as depicted in Figure \ref{fig1}. According to the inverse pole figure (IPF) color coding, the deformed grains are aligned with the \hkl[10-10] axis parallel to ED, while the recrystallized grains exhibit random orientations. This indicates a texture modification effect induced by dynamic recrystallization (DRX) and metadynamic recrystallization during hot extrusion and air cooling. 

Site-specific preparation of an APT tip lifted out from a deformed region in Figure \ref{fig1} was performed by a coordinated process of focused ion beam (FIB) milling and TKD in a FEI Helios 600i dual-beam electron microscope to ensure a precise position of the examined LAGB within the APT tip. Details of this method can be found in \cite{pei2023atomistic}. The elemental distributions of solute atoms in the vicinity of the measured LAGB was characterized by 3D-APT in a Local Electrode Atom Probe 4000X HR from Cameca using laser-pulsing mode at a temperature of 30 K. Reconstruction of the evaporated tips was performed using the software package IVAS 3.8.2. 

To evaluate the clustering tendency of solute elements in the APT reconstructed sample, we performed a cluster analysis implemented in OVITO \cite{stukowski2009visualization} using a cutoff distance of 5 \AA. The solute clusters containing more than one atom and their compositions were characterized for the GB region and in the matrix.

\subsection{Density functional theory calculations}
The binding energies of solute pairs in the Mg matrix were calculated using DFT. The calculations followed the methodology outlined in \cite{mouhib2024exploring} using the Quantum ESPRESSO package \cite{giannozzi2009quantum}, employing the projector augmented wave (PAW) method \cite{kresse1999ultrasoft,blochl1994projector} and the Perdew-Burke-Ernzerhof (PBE) generalized gradient approximation (GGA) \cite{perdew1996generalized} for the exchange-correlation functional. A kinetic energy cutoff of 40 Ry was set for wave functions, while a cutoff of 280 Ry was used for charge densities and potentials. Electronic self-consistency was achieved with a convergence threshold of 10$^{-8}$ Ry. Structural relaxations were conducted using the Broyden-Fletcher-Goldfarb-Shanno (BFGS) relaxation scheme \cite{fletcher2000practical}, with energy and force convergence thresholds set at 10$^{-4}$ a.u. and 2 $\times$ 10$^{-4}$ a.u., respectively. The orthorhombic simulation cells comprised 5 $\times$ 3 $\times$ 3 unit cells with 180 atoms using a 4 $\times$ 4 $\times$ 4 k-point mesh. Solute-solute binding energies were calculated using the following equations:
\begin{equation}
\label{equ1}
\Delta E_{\text{bind}}^\text{b,I-J} = (E_{\text{b}}^\text{I-J}+E_{\text{b}}) - (E_{\text{b}}^\text{I} + E_{\text{b}}^\text{J})
\end{equation}
where $E_\text{b}$ is the energy of the Mg matrix, $E_{\text{b}}^{I}$ and $E_{\text{b}}^{J}$ are the energies of the Mg matrix containing I and J solutes, respectively, and $E_{\text{b}}^{I-J}$ is the energy of the Mg matrix with a solute pair of I and J atoms. Here, a negative binding energy $\Delta E_{\text{bind}}^\text{b,I-J}$ indicates an attractive solute interaction, i.e. favorable binding.

\subsection{Atomistic simulations}
Atomistic simulations were performed using the open-source MD software package LAMMPS \cite{thompson2022lammps}, employing modified embedded atom method (MEAM) potentials for Mg-Ca-Zn \cite{jang2019modified}, Mg-Ca-Al, and Mg-Zn-Al \cite{jang2021modified} with identical pair interactions. The potential properties of pure Mg, including elastic constants, stacking fault energies, and grain boundary energies, agree well with the experimental and ab-initio values \cite{wang2024defects}. Additionally, per-site segregation energies of Ca, Zn and Al solutes at $\Sigma$ 7 21.8° \hkl{12-30}\hkl<0001> grain boundary, \hkl{10-11} and \hkl{10-12} twin boundaries aligned well with the ab-initio calculations \cite{wang2024defects}. The binding energies of solute pairs in the Mg matrix, computed using the same 5 $\times$ 3 $\times$ 3 unit cells as in DFT calculations. 

\begin{figure*}[ht!]
\centering
\includegraphics[width=\textwidth]{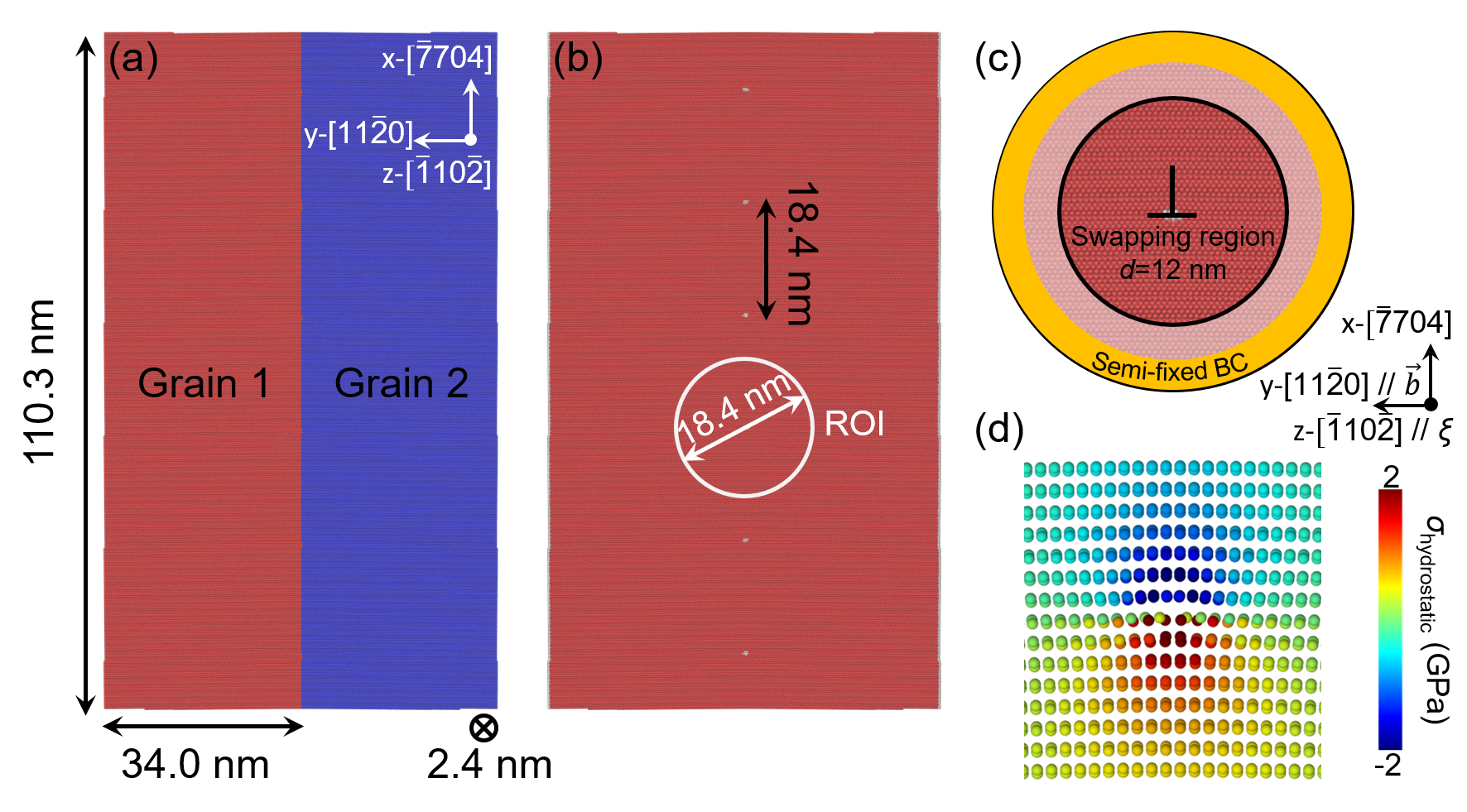}
\caption{Atomistic configuration of the symmetric tilt LAGB with a misorientation of 1° along the \hkl[-110-2] rotation axis on the \hkl(11-20) GB plane. (a) Slab setup (110.3 × 68.0 × 2.4 nm$^{3}$) of the symmetric tilt LAGB with periodic boundary conditions in x and z directions. (b) The relaxed symmetric tilt LAGB, consisting of an array of edge dislocations with identical core structures and a spacing of 18.4 nm. Atoms are colored according to the common neighbor analysis \cite{honeycutt1987molecular}, with white indicating atoms at dislocation cores and red for those in the matrix. (c) Schematic of the cylindrical setup ($d$=18.4 nm, $l_{z}$=2.4 nm) used for solute segregation calculation at the dislocation, featuring semi-fixed boundary conditions (constrained in x and y directions) at outermost layers with a thickness of 1.4 nm. (d) Hydrostatic stress map illustrating the stress fields around the dislocation core region, where blue and red regions represent compressive and tensile stress fields of the edge dislocation, respectively. 
}
\label{fig2}
\end{figure*}

The atomistic configuration of the symmetric tilt LAGB was constructed using Atomsk \cite{hirel2015atomsk}, informed by experimentally determined crystallographic orientations of the grains and GB plane normal to the y-axis. The orientation was obtained through TKD grain mapping and reconstructed APT tip (shown in Figure \ref{fig4}). The simulation setup is illustrated in Figure \ref{fig2}. Microscopic degrees of freedom of the LAGB were explored by performing rigid body shift and structural relaxation normal to the GB plane with a force tolerance of $10^{-8}~\text{eV/\AA}$ using the FIRE algorithm \cite{bitzek2006structural,guenole2020assessment}. A 2D cylindrical region with a radius of 18.4 nm and a dimension of 2.4 nm in the periodic direction was cut out in the fully relaxed sample to investigate the segregation behavior at the dislocation array, as shown in Figure \ref{fig2} b-c. The outermost layers of the cylindrical surface, with a thickness of 1.2 nm (equivalent to twice the interatomic potential cutoff), were constrained in the x and y directions. Periodic boundary conditions were applied along the dislocation line in the z-direction. The boundary effect on calculated per-site segregation energy was found to be negligible, being less than 0.1 meV. Atomistic configurations were visualized using OVITO. 

\subsection{Thermodynamic models of GB segregation}

\begin{figure*}[ht!]
\centering
\includegraphics[width=\textwidth]{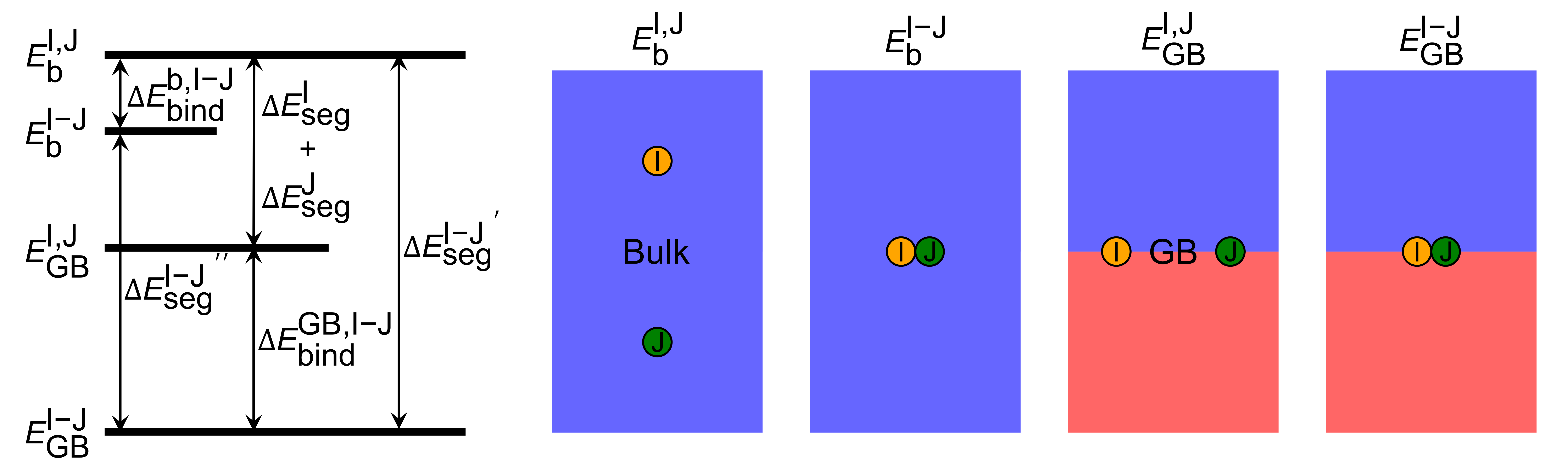}
\caption{Schematic illustration of the relationships between energy states ($E$), segregation energies ($\Delta E_\text{seg}$), co-segregation energies ($\Delta E_\text{seg}^\text{I-J}$), and binding energies ($\Delta E_\text{bind}^\text{I-J}$) for solutes I and J in bulk Mg and at the GB.}
\label{fig3}
\end{figure*}

Thermodynamic models for GB segregation were applied using atomistic simulation results. After substituting a matrix atom with a solute atom at site $i$, energy minimization using the FIRE algorithm was conducted. The per-site segregation energy $\Delta E_{\text{seg},i}$ was calculated as:
\begin{equation}
\label{equ2}
\Delta E_{\text{seg},i} = (E_\text{GB}^i - E_\text{GB}) - (E_\text{b}^i-E_\text{b}),
\end{equation}
where $E_\text{b}^i$ is the energy of the Mg matrix with one host atom replaced by a solute atom, $E_\text{GB}$ is the energy of a system with a GB, and $E_\text{GB}^i$ is the energy of the system with a solute atom occupying a GB site or a site in the dislocation core region. In the above-mentioned definition, a more negative value means a higher tendency for segregation. 

Figure~\ref{fig3} illustrates the energy states ($E$) and energy differences ($\Delta E$) for various solute interaction scenarios within the matrix and at the GB \cite{elsasser2005codoping,scheiber2018impact}. Four possible configurations are depicted: (i) Two separate solute atoms, I, J, in the bulk corresponding to an energy state $E_\text{b}^\text{I,J}$, (ii) a solute pair, I-J, in the bulk with $E_\text{b}^\text{I-J}$, (iii) two separate solute atoms, I and J, at the GB with $E_\text{GB}^\text{I,J}$, and (iv) a solute pair, I-J, at the GB represented by $E_\text{GB}^\text{I-J}$. The binding and (co-)segregation energies were calculated to assess the tendency for forming solute pairs or clusters.

The mono-segregation energy $\Delta E_\text{seg}^{\text{I-J}\prime}$ is calculated as:
\begin{equation}
\label{equ3}
\Delta E_\text{seg}^{\text{I-J}\prime} = (E_\text{GB}^\text{I-J} - E_\text{GB}) - (E_\text{b}^\text{I,J} - E_\text{b}),
\end{equation}
where $\Delta E_\text{seg}^{\text{I-J}\prime}$ describes a mono-segregation scenario, where separate $I $ and $J$ solute atoms in the bulk individually segregate to the GB, form a solute pair I-J that occupies two adjacent GB sites (either direct first neighbors or second nearest neighbors). A negative value of $\Delta E_\text{seg}^{\text{I-J}\prime}$ indicates that forming a solute pair at the GB is energetically more favorable for the solute pair to reside at the GB than to remain as a pair in the bulk.

Analogously, the co-segregation energy $\Delta E_\text{seg}^{\text{I-J}\prime\prime}$ is defined as:
\begin{equation}
\label{equ4}
\Delta E_\text{seg}^{\text{I-J}\prime\prime} = (E_\text{GB}^\text{I-J} - E_\text{GB})-(E_\text{b}^\text{I-J} - E_\text{b}),
\end{equation}
where $\Delta E_\text{seg}^{\text{I-J}\prime\prime}$ describes another segregation scenario (co-segregation). In this context, a solute pair I-J in the bulk migrates together to form a cluster at the GB, occupying two adjacent GB sites. A negative value of $\Delta E_\text{seg}^{\text{I-J}\prime\prime}$ indicates that it is  the solute pair at the GB is energetically preferred over the bulk pair.

Finally, the energetic advantage of solute binding at the GB over a separate distribution at distant GB sites is expressed as $\Delta E_\text{bind}^{\text{GB,I-J}}$, following a similar definition as described in Equation~\eqref{equ1}. A negative value of $\Delta E_\text{bind}^{\text{GB,I-J}}$ forming solute pairs at the GB is more energetically favorable than having them distributed separately across the GB. 

The solute concentration at GBs is directly related to the per-site segregation energy, as expressed in the spectral segregation model of the Langmuir–McLean isotherm \cite{mclean1958grain,white1977spectrum}. For an infinitesimally dilute solid solution, this is represented by: 
\begin{equation}
\label{equ5}
X_{\text{GB},i} = \left(1 + \frac{1 - X_{\text{b}}}{X_{\text{b}}} \exp\left(\frac{\Delta E_{\text{seg},i}}{k_\text{B} T}\right)\right)^{-1},
\end{equation}
where $X_{\text{b}}$ is the bulk solute concentration, and $X_{\text{GB},i}$ is the solute concentration at a specific GB site $i$. This formulation facilitates the calculation of the weighted average solute concentration at GBs ($X_\text{GB}$), given by: 
\begin{equation}
\label{equ6}
X_\text{GB} = \sum_i F_i X_{\text{GB},i},
\end{equation}
where $F_i$ is the fraction of GB sites associated with $X_{\text{GB},i}$. This approach directly correlates theoretical predictions from atomic-scale modeling with experimentally measured solute concentrations at GBs through 3D-APT.

Building on the Langmuir–McLean framework, the Guttmann model \cite{guttmann1975equilibrium,guttmann1979grain} incorporates solute-solute interactions within the GB region through solute pair interactions, characterized by an interaction parameter $\Omega$. This modification accounts for deviations from ideal segregation behavior, particularly in systems that exceed dilute limits, where solute clustering or repulsion becomes significant. The segregation energy is enhanced with a pair interaction term, $\Delta E_{\text{seg},i}^{\text{pair}}$, expressed as:
\begin{equation}
\begin{split}
\label{equ7}
\Delta E_{\text{seg},i}^{\text{pair}} = -2\left( \Omega_{\text{GB}}^\text{I-M} X_{\text{GB}}^\text{I} - \Omega_{\text{b}}^\text{I-M} X_{\text{b}}^\text{I}\right)\\+ \sum_{I \neq J} \left(\Omega_{\text{GB}}^\text{'I-J} X_{\text{GB}}^\text{J} - \Omega_{\text{b}}^\text{'I-J} X_{\text{b}}^\text{J}\right),
\end{split}
\end{equation}
Here, $\Omega^\text{I-J}$ describes the interaction of solutes I and J at the sites $i$ and $j$, respectively. It depends on the coordination number $Z$ of the site $i$ and the differences in bond energies of solute pairs: 
\begin{equation}
\begin{split}
\label{equ8}
\Omega^\text{I-J} = \frac{1}{2}Z\omega^\text{I-J},
\end{split}
\end{equation}
where $\omega^\text{I-J}=\Delta E^\text{I-J} - \left(\Delta E^\text{I-I} + \Delta E^\text{J-J}\right)/2$.

The interaction coefficient $\Omega^\text{'I-J}$ represents the net interaction between solutes I and J with respect to their interactions with the matrix element M: 
\begin{equation}
\label{equ9}
\begin{split}
\Omega^\text{'I-J} = \Omega^\text{I-J} - \Omega^\text{I-M} - \Omega^\text{J-M}.
\end{split}
\end{equation}
The first term in Equation~\eqref{equ7} represents the Fowler interaction \cite{fowler1939statistical}, which only considers I-I or J-J binary interactions. The effective segregation energy becomes:
\begin{equation}
\label{equ11}
\Delta E_{\text{seg},i}^{\text{eff}} = \Delta E_{\text{seg},i} + \Delta E_{\text{seg},i}^{\text{pair}},
\end{equation}
The spectral segregation model is thus extended as per Equation~\eqref{equ11} to include interaction effects. Incorporating solute-solute interaction effects allows the model to better capture the co-segregation behavior at higher solute concentrations in the GB region, where such interactions become significant.

\section{Results}
\subsection{3D-APT characterization of solute segregation to LAGB}
\begin{figure*}[ht!]
\centering
\includegraphics[width=\textwidth]{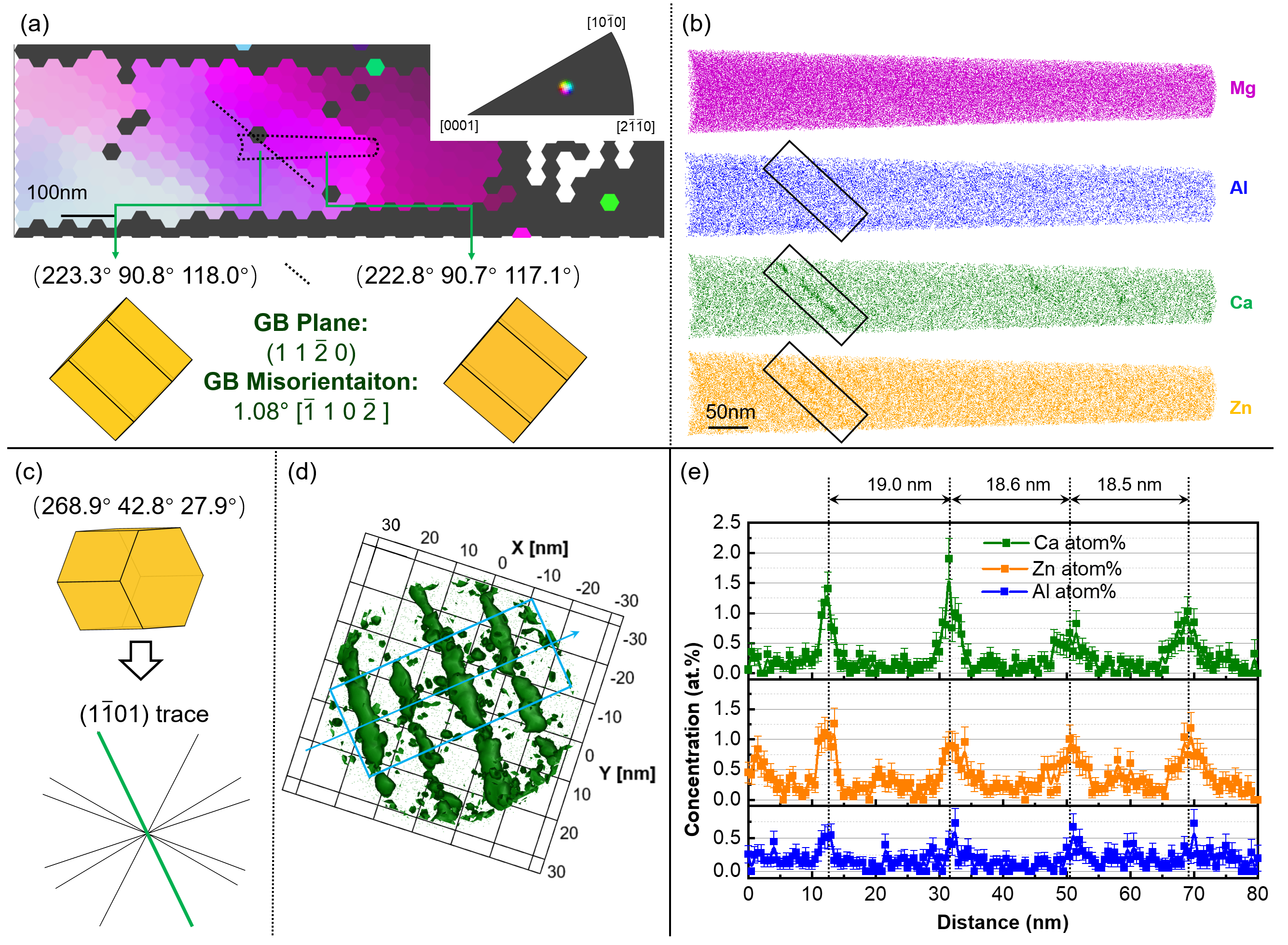}
\caption{Chemical analysis of the LAGB in the APT specimen of deformed AZX010 Mg alloy. (a) TKD grain map of the APT specimen, with the IPF color key restricted to a 5° deviation from the average orientation. (b) Elemental distribution maps for Mg, Al, Ca and Zn within the reconstructed volume, highlighting solute segregation at the LAGB in the outlined region. (c) Corresponding grain orientation viewed from the top of the tip and the \hkl{1-101} plane traces calculated based on Euler angles obtained from TKD. (d) 1.0 at.\% Ca isosurface (X-Y plane) within the region of interest marked in (b), indicating parallel segregation patterns along line defects in the LAGB. (e) Solute concentration profiles (at.\%) across the segregated regions, measured within a cylindrical volume 30 nm in diameter. The profiles reveal periodic segregation with an approximate spacing of 19 nm.}
\label{fig4}
\end{figure*}

For a high-resolution crystallographic analysis of the investigated LAGB, Figure \ref{fig4}a presents a TKD map of the APT specimen prior to the last annular milling step. To illustrate the lattice distortion within the deformed grain, the IPF color key in the map was limited to a 5° deviation from the average orientation across the entire map. The position of the final needle-shaped volume containing the LAGB is depicted in Figure \ref{fig4}a, located within 200 nm from the apex. The grain orientations derived from TKD revealed a GB misorientation of 1.08° along \hkl[-110-2], with the GB plane being \hkl(11-20). Figure \ref{fig4}b presents elemental distribution maps for Mg, Ca, Zn and Al atoms within the reconstructed APT tip, indicating solute enrichment of Ca, Zn and Al atoms at the observed LAGB. The morphology of segregated solute regions associated with the LAGB is distinctly visible from a top view of the tip, as illustrated in Figure \ref{fig4}d. This view reveals parallel linear solute decorations (1.0~at.\% Ca iso-surfaces) in the X-Y plane running perpendicular to the major axis of an analyzed cylindrical volume. 

The crystallographic analysis shown in Figure \ref{fig4}c reveals that regions where Ca has segregated lie on the pyramidal \hkl(1-10-1) plane. Figure \ref{fig4}e provides concentration profiles (at \%) of Ca, Zn and Al across these parallel segregated regions within a 30 nm-diameter cylindrical volume outlined in Figure \ref{fig4}d. The segregation pattern observed in Figure \ref{fig4}d is interpreted as representing an array of Read–Shockley dislocations lying on the LAGB plane \cite{read1950dislocation}. The quantitative concentration profiles, particularly of Ca, suggest that the segregated region is about 3 nm wide and indicate a spacing of $18.7 \pm 0.2$~nm between dislocations, as shown in Figure \ref{fig4}e. The derived misorientation angle from TKD aligns well with the theoretical value determined from the Frank-Bilby formulation~\cite{frank1950resultant,bilby1955bristol}, based on the Burgers vector ($b = 0.32~\text{nm}$) and the distance between edge dislocations in a pure tilt GB:
\begin{equation}
\label{equ12}
\theta = 2 \arcsin\left(\frac{b}{2d}\right) = 0.98 \pm 0.2^\circ
\end{equation}
Additionally, it is evident that there are variations in segregation levels at each dislocation position within the array plane. These variations might be attributed to defects of atomic dimensions, such as kinks or jogs along the dislocation line.

\begin{figure*}[ht!]
\centering
\includegraphics[width=\textwidth]{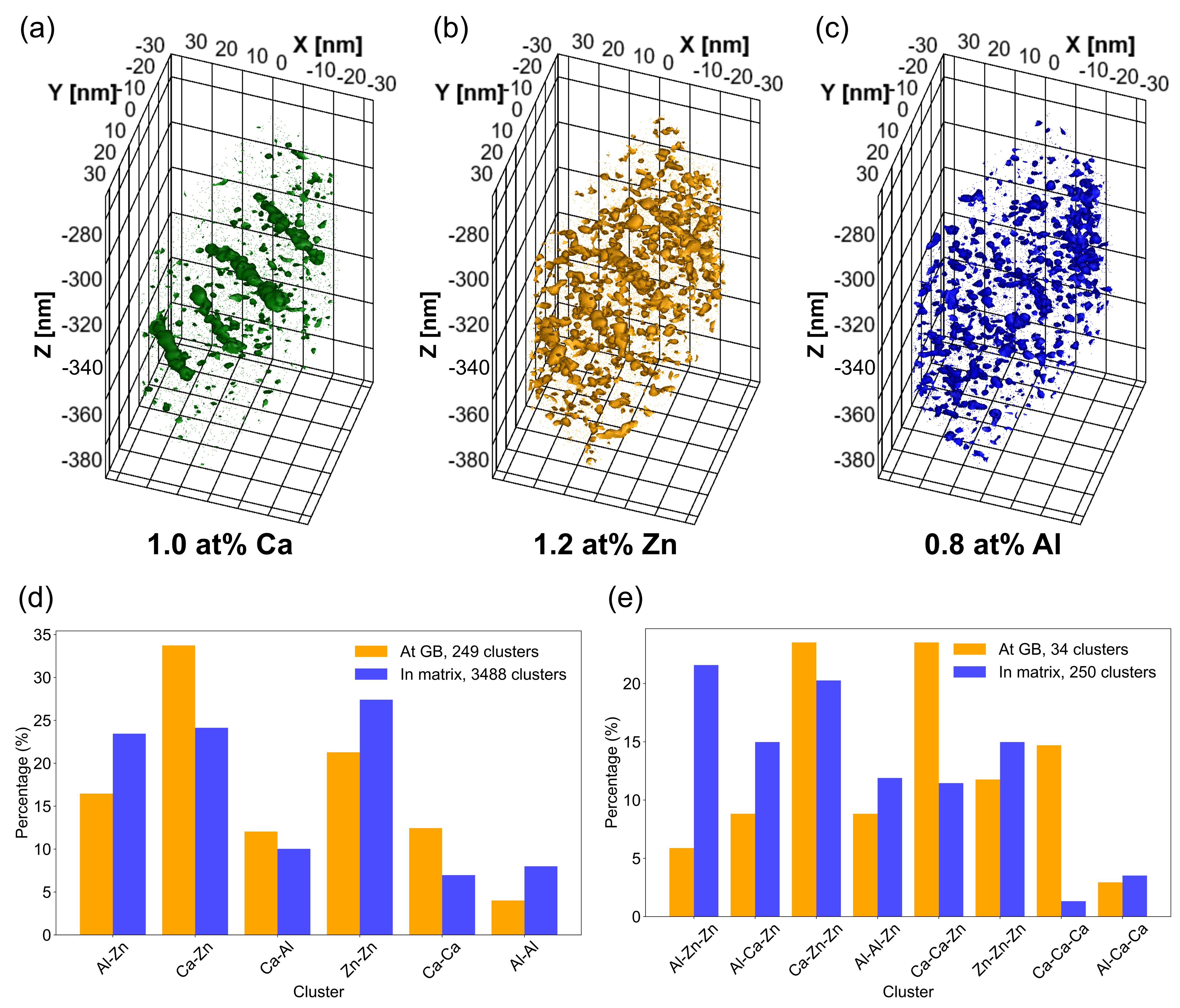}
\caption{3D reconstructions of the isosurfaces in the GB region for Ca (a), Zn (b) and Al (c). The revealed segregation morphology is closely associated with the line defects constituting the investigated LAGB. Clustering analysis of all solute pairs in the APT dataset (d) and the most frequent solute triplets (clusters), both in the matrix and at the GB region, defined by cylindrical areas along the segregated linear regions with a radius of 6 nm.}
\label{fig5}
\end{figure*}

The average peak concentrations of Ca, Zn, and Al within the dislocation array are listed in Table \ref{table1}. The measured Ca, Zn, and Al average peak concentrations at the LAGB were 1.29, 1.10, and 0.70~at.\%, respectively. Notably, Ca atoms demonstrated a significantly stronger segregation potential with a segregation ratio of 12.56, whereas Zn and Al exhibited similar potentials with segregation ratios of approximately 4.05 and 4.65, respectively. This result is further illustrated in the isosurface maps shown in Figure \ref{fig5}a-c, where thresholds were set at isoconcentrations of 1.0 at.\% for Ca, 1.2 at.\% for Zn and 0.8 at.\% for Al. 

\begin{table*}[ht!]
    \centering
    \caption{Solute concentrations of Ca, Zn and Al within the LAGB obtained from the concentration profiles in Figure \ref{fig4}d . The positions 1-4 denote the periodic locations along the boundary that are intersected with the array dislocations. Peak concentrations denote the maximum solute concentration measured at the disloaction. Baseline concentrations denote the concentration level away from the peaks, i.e. in the absence of dislocations. SD: Standard deviation.}
    \begin{tabular}{l|cccccc}
    \hline
    
    Position & \multicolumn{2}{c}{Ca} & \multicolumn{2}{c}{Zn} & \multicolumn{2}{c}{Al} \\

    & Peak& Baseline& Peak& Baseline& Peak& Baseline\\ 
    \hline
    
     \# 1     & 1.43  & 0.14      & 1.27  & 0.22      & 0.59  & 0.16      \\ 
     \# 2     & 1.92  & 0.11      & 0.95  & 0.30      & 0.76  & 0.15      \\ 
     \# 3     & 0.78  & 0.06      & 1.02  & 0.22      & 0.68  & 0.14      \\ 
     \# 4     & 1.02  & 0.10      & 1.14  & 0.34      & 0.76  & 0.15       \\
     Mean     & 1.29  & 0.10      & 1.10  & 0.27      & 0.70  & 0.15       \\
     SD& 0.39  & 0.02      & 0.11  & 0.05      & 0.06  & 0.005       \\
     \hline
     Segregation ratio (Peak/Baseline)& \multicolumn{2}{c}{12.90} & \multicolumn{2}{c}{4.07} & \multicolumn{2}{c}{4.67} \\
     \hline
  \end{tabular}
  \label{table1}
\end{table*}

The clustering analysis of solute elements in the APT-reconstructed sample revealed interesting patterns, as illustrated in Figure \ref{fig5}(d,e). The data presents statistics for solute pairs or small clusters containing three solute atoms within a cutoff distance of 5 Å, both at the GB and in the matrix. The GB region is defined by cylindrical areas along the segregated dislocation lines with a radius of 6 nm, while the matrix is identified as the area beyond 10 nm from these segregation lines. Among solute pairs, Zn-Zn pairs exhibit the highest fraction in the matrix, followed by Ca-Zn and Al-Zn pairs, with Ca-Ca pairs being the least frequent. However, near the GB, there is a notable increase in Ca-Ca pair fractions from 7\% to 12.4\%. Other Ca-containing pairs also show an increase near the GB, especially Ca-Zn pairs, which rise from 24\% to 34\%. In contrast, Al-Zn, Zn-Zn, and Al-Al pairs decrease in prevalence within the GB region. For solute clusters, Al-Zn-Zn and Ca-Zn-Zn dominate in the matrix, followed by Ca-Zn-Zn and Zn-Zn-Zn. Near the GB, Ca-Ca-Ca clusters increase significantly from 1.2\% to 14.7\%. Other Ca-containing clusters, such as Ca-Zn-Zn and Ca-Ca-Zn, also increase, with Ca-Ca-Zn clusters rising from 10.4\% to 23.5\%. Conversely, clusters containing Al and Zn decrease near the GB, most notably Al-Zn-Zn, which drop from 19.6\% to 5.9\%.

\subsection{Cross-scale modeling: Mono-segregation to dislocation array in LAGB}
To gain further insights into the atomistic mechanisms of solute clustering and segregation at the LAGB, atomistic simulations for dislocation arrays with experimentally informed characteristics were performed.

\begin{figure*}[ht!]
\centering
\includegraphics[width=\textwidth]{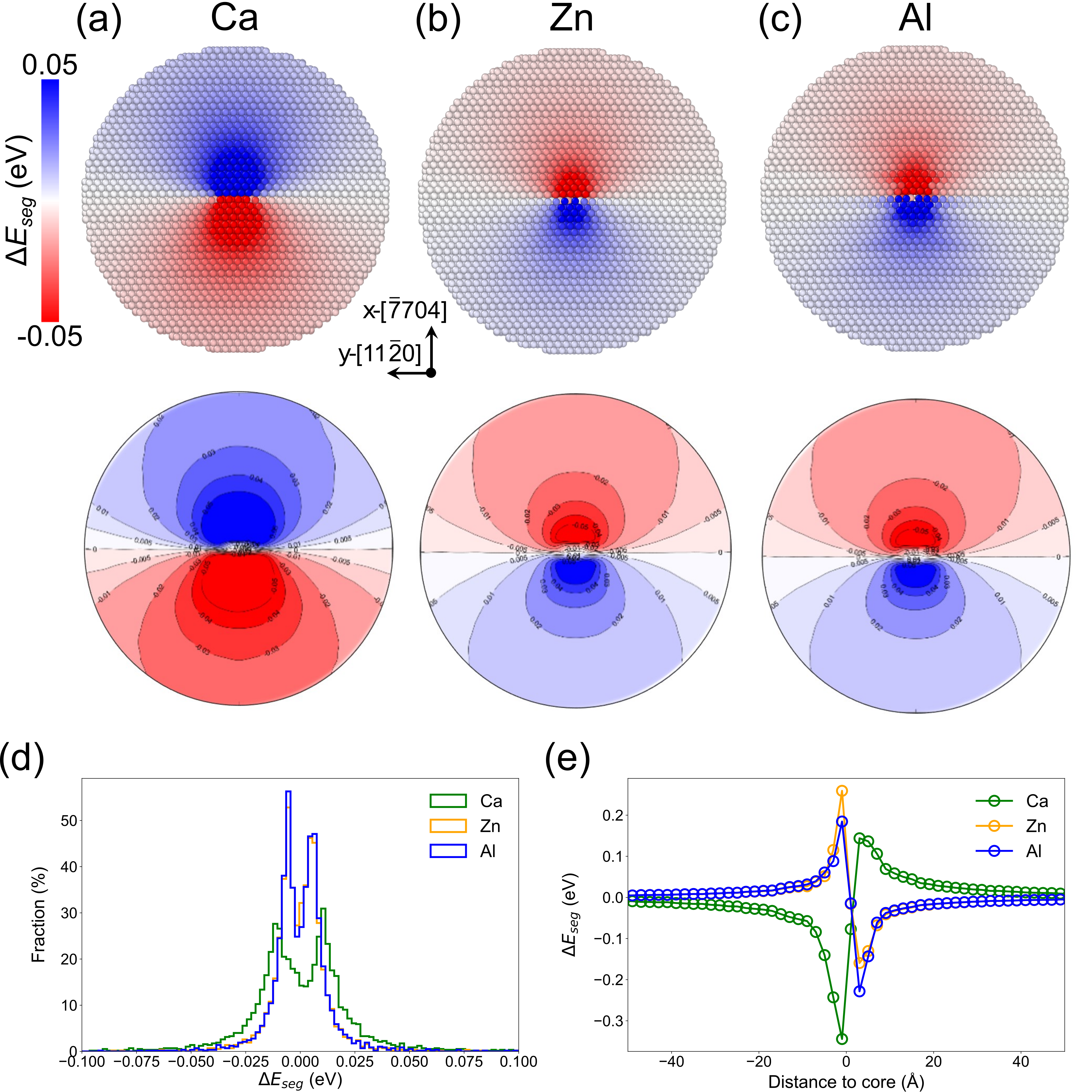}
\caption{Mono-segregation behavior of (a) Ca, (b) Zn and (c) Al solutes at the dislocation. Upper row: Atomistic simulations; lower row: Interaction energy fields between the solute modeled as an elastic dipole and the strain field of the dislocation. A negative value of $\Delta E_\text{seg}$ indicates that segregation is energetically favorable. (d) Distribution of $\Delta E_\text{seg}$ for Ca, Zn and Al solutes with data grouped into bins of 2 meV. (e) Statistics of $\Delta E_\text{seg}$ as a function of distance to the center of the dislocation core. Negative distances correspond to the tensile stress region, whereas positive distances indicate the compressive stress region. Data is divided into bins of 2 \AA.}
\label{fig6}
\end{figure*}

\begin{figure*}[ht!]
\centering
\includegraphics[width=0.9\textwidth]{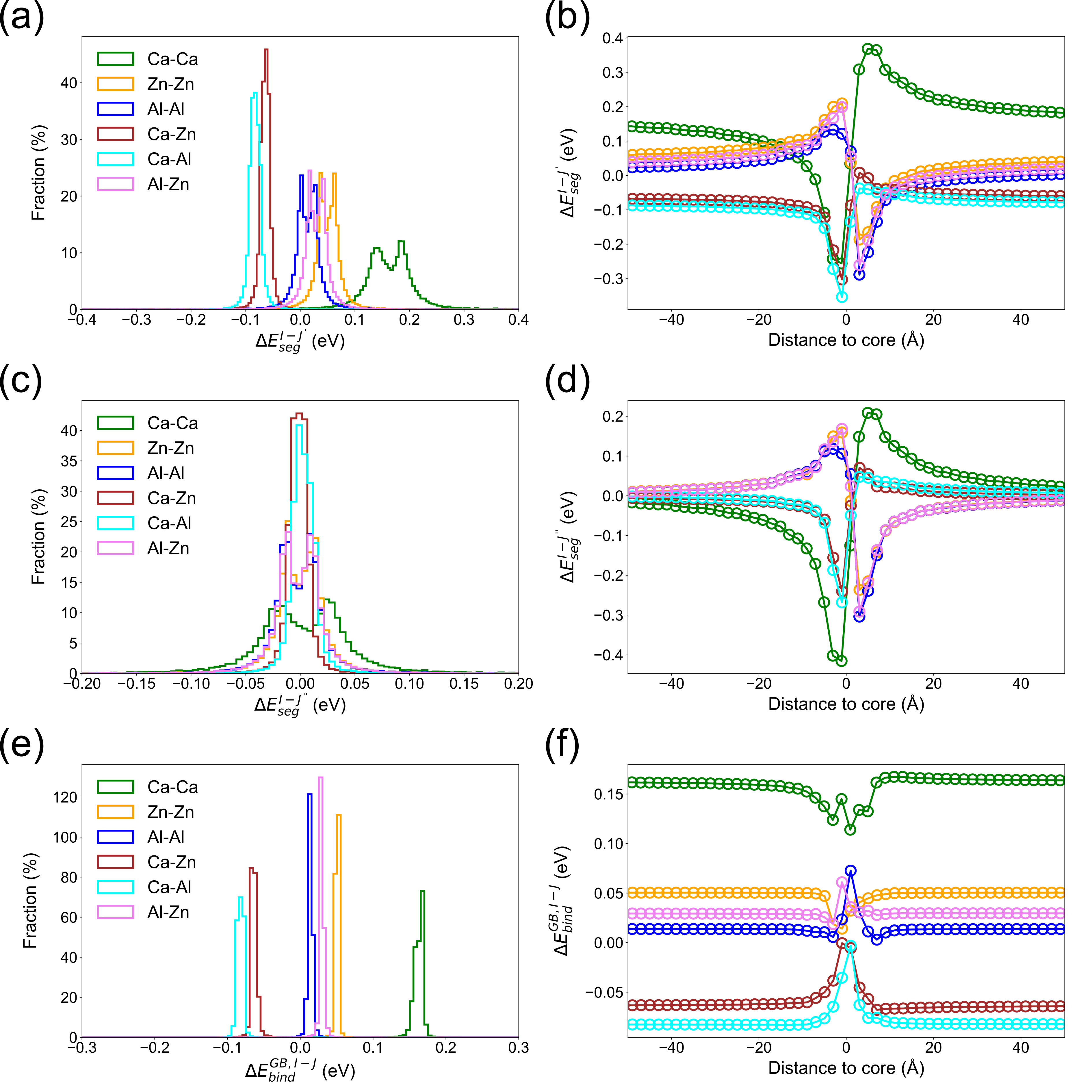}
\caption{Atomistic simulations of co-segregation behavior of Ca, Zn and Al solutes at the dislocation. (a) Distribution of $\Delta E_\text{seg}^{\text{I-J}\prime}$, and (b) statistics of $\Delta E_\text{seg}^{\text{I-J}\prime}$ as a function of distance to the center of the dislocation core. (c) Distribution of $\Delta E_\text{seg}^{\text{I-J}\prime\prime}$, and (d) statistics of $\Delta E_\text{seg}^{\text{I-J}\prime\prime}$ as a function of distance to the center of the dislocation core. (e) Distribution of $\Delta E_\text{bind}^{\text{GB,I-J}}$, and (f) statistics of $\Delta E_\text{bind}^{\text{GB,I-J}}$ as a function of distance to the center of the dislocation core. Negative distances correspond to the tensile stress region, whereas positive distances indicate the compressive stress region. The bin size in (a), (c) and (e) is 5 meV, and in (b), (d), (f) 2 \AA.}
\label{fig7}
\end{figure*}

Figure \ref{fig6}a-c shows the distribution of per-site segregation energy $\Delta E_\text{seg}$ of segregated Ca, Zn and Al solutes to a model edge dislocation. As depicted in Figure \ref{fig2}d, Ca solutes exhibit a preference for segregation in the tensile stress field. Conversely, Zn and Al solutes tend to segregate in the compressive stress field. Accordingly, all $\Delta E_\text{seg}$ spectra in Figure \ref{fig6}d exhibit a bimodal distribution, reflecting the distinct segregation preferences between tensile and compressive stress fields. The $\Delta E_\text{seg}$ distribution for Ca displays a broader peak compared to those of Zn and Al solutes. This difference become more apparent when $\Delta E_\text{seg}$ is plotted against the distance from the dislocation core, as shown in Figure \ref{fig6}e. The $\Delta E_\text{seg}$ values for Ca consistently remain higher in the tensile stress field compared to $\Delta E_\text{seg}$ values for Zn and Al in the compressive stress field. Near the dislocation core, Al shows a slightly stronger segregation tendency than Zn, as evidenced by its more negative $\Delta E_\text{seg}$ values. The simulation results on mono-segregation align well with the experimental observation of solute enrichment at the LAGB, where Ca exhibits a segregation ratio twice that of Zn and Al.

In addition to the segregation energy computed by atomistic simulations, permanent elastic dipole tensors offer valuable insights into the observed mono-segregation behavior within elasticity theory. Elastic dipole calculations, employed as a framework for modeling point defects, correspond to the first moment of the distribution of point forces imposed by a solute atom on its surrounding atoms, reflecting size and shape effects \cite{LOV1959, KRO1959, SIE1968, LEI1978, BAL2016, CLO2018, GEN2024}. These tensors can be determined through molecular statics simulations based on the residual stress tensor of a bulk lattice containing a single solute~\cite{CLO2018, PET2025}. The permanent elastic dipole tensors $P_{ij}^0$ of a single Ca, Zn, and Al solute atom in substitution in the Mg lattice were computed and are presented as follows in the frame $X \parallel [\overline{7}704]$, $Y \parallel [11\overline{2}0]$, $Z \parallel [\overline{1}10\overline{2}]$:
\[
\text{Ca in Mg:} \quad P_{ij}^0 = 
\begin{pmatrix}
3.19 & 0 & 0.03 \\
0 & 3.18 & 0 \\
0.03 & 0 & 3.24
\end{pmatrix} \, \text{(eV)}
\]

\[
\text{Zn in Mg:} \quad P_{ij}^0 = 
\begin{pmatrix}
-1.70 & 0 & -0.03 \\
0 & -1.69 & 0 \\
-0.03 & 0 & -1.74
\end{pmatrix} \, \text{(eV)}
\]

\[
\text{Al in Mg:} \quad P_{ij}^0 = 
\begin{pmatrix}
-1.63 & 0 & 0.09 \\
0 & -1.68 & 0 \\
0.09 & 0 & -1.52
\end{pmatrix} \, \text{(eV)}
\]  

The solute segregation preferences observed in atomistic simulations align well with the elastic dipole values. Both Al and Zn exhibit negative principal values that are similar in  magnitude, consistent with their comparable segregation field amplitudes in Mg. This consistency extends to Ca, which displays a stronger segregation field intensity with an opposite sign. The elastic dipole values of Ca are indeed positive and higher in magnitude compared to Al and Zn. Additionally, elastic dipole tensors enable the calculation of the interaction energy between a solute atom and the strain field $\varepsilon_{ij}$ generated by crystalline defects such as dislocations ~\cite{KRO1959,DEM2021,CLO2018}, using :
\begin{equation}
E^\text{int}(x)=-P_{ij}\varepsilon_{ij}(x)
\end{equation}
The strain field around the dislocation is computed through molecular statics simulations and subsequently interpolated onto a regular grid ~\cite{TOR2015,PET2025}. Figure \ref{fig6} then illustrates the interaction energy fields of Ca, Zn and Al solutes as they interact with the dislocation, demonstrating good accuracy of the interaction energy model in capturing the mono-segregation behavior of studied solutes.

\subsection{Atomic-scale modeling: Solute binding and co-segregation to dislocation array in LAGB}

\begin{figure}[ht!]
\centering
\includegraphics[width=0.45\textwidth]{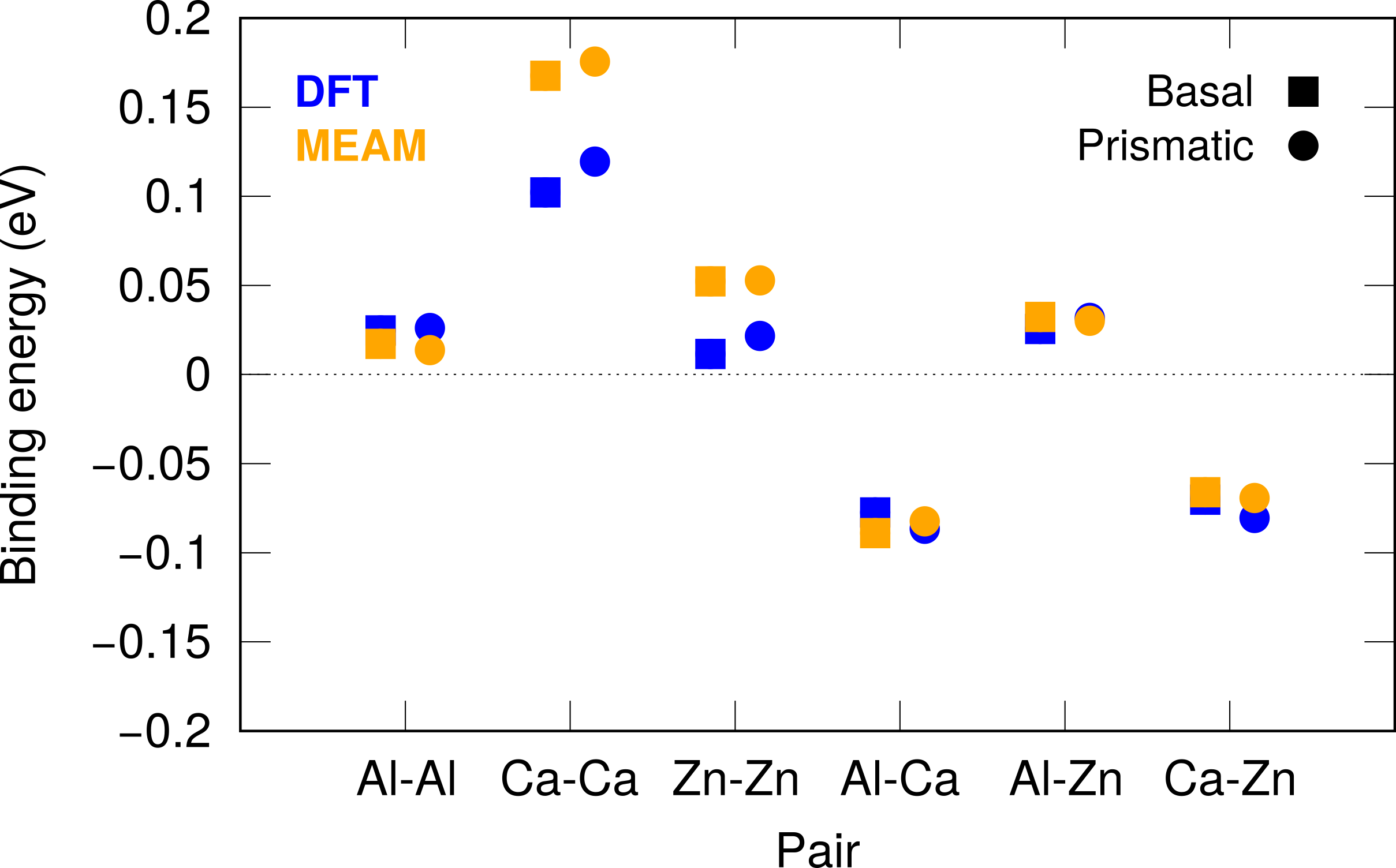}
\caption{Binding energies of solute pairs in the Mg matrix (5 $\times$ 3 $\times$ 3 unit cells) calculated using the MEAM potentials and DFT.}
\label{fig8}
\end{figure}

To gain a deeper understanding of the strong clustering tendency of solutes observed in post-processed APT data (Figure \ref{fig5}(d-e)), the binding energy between different solute pairs at dislocations in ternary systems (Mg-Al-Ca, Mg-Al-Zn, Mg-Zn-Ca) was modeled using atomistic simulations. The analysis focused on solute pairs within the cutoff distance of the first nearest neighbor. The obtained values showed strong agreement with binding energy results derived from DFT, as illustrated in Figure \ref{fig8}. 

Returning to Figure \ref{fig3} and Equations \ref{equ3} and \ref{equ4}, the distribution of segregation energies in the ternary systems was modeled on the basis of different clustering and segregation scenarios from the bulk to the dislocation core region at the LAGB, as depicted in Figure \ref{fig7}.  One scenario involves separate solute atoms in the bulk clustering individually to form a solute pair at the dislocation, as described by Equation~\eqref{equ3}. Figure \ref{fig7}a presents the corresponding distribution of segregation energy $\Delta E_\text{seg}^{\text{I-J}\prime}$ for various solute pairs within a 6 nm radius of the dislocation core. Notably, Ca-Zn and Ca-Al solutes predominantly exhibit negative $\Delta E_\text{seg}^{\text{I-J}\prime}$ values with narrow distributions, indicating favorable segregation and clustering at the dislocation region. In contrast, Ca-Ca solutes mostly show positive $\Delta E_\text{seg}^{\text{I-J}\prime}$ values with a bi-modal distribution. Similarly, Zn-Zn, Al-Zn, and Al-Al solutes also exhibit bi-modal distributions, with most values leaning towards the positive side of the distribution. Figure \ref{fig7}(b) shows the distribution of $\Delta E_\text{seg}^{\text{I-J}\prime}$ as a function of distance to the dislocation core, providing insights into spatial segregation tendencies of solute pairs. For Ca-Ca solutes, although most regions near the dislocation core are unfavorable for segregation, the tensile stress field within approximately 8 Å exhibits negative $\Delta E_\text{seg}^{\text{I-J}\prime}$ values, indicating favorable Ca-Ca segregation. Similarly, other Ca-containing pairs like Ca-Zn and Ca-Al demonstrate stronger segregation tendencies in the tensile stress field. In contrast, Zn-Zn, Al-Zn, and Al-Al pairs exhibit more favorable segregation in the compressive stress field of the dislocation core.

An alternative scenario of solute segregation involves solute pairing within the bulk, followed by co-segregation to the dislocation core region, as described by Equation~\eqref{equ4} and illustrated in Figures \ref{fig7}c and d. In this case, the distributions of $\Delta E_\text{seg}^{\text{I-J}\prime\prime}$ for all solute pairs are centered around zero, with Ca-Ca pairs displaying a broader distribution compared to others. Conversely, Ca-Zn and Ca-Al pairs exhibit the narrowest distributions, while the remaining pairs display bi-modal distributions. Similar to the first scenario, Ca-containing pairs (Ca-Zn, Ca-Al, and Ca-Ca) preferentially co-segregate into the tensile stress field. Among these, Ca-Ca pairs show the strongest preference, as indicated by larger negative $\Delta E_\text{seg}^{\text{I-J}\prime\prime}$ values. In contrast, Zn-Zn, Al-Zn, and Al-Al pairs demonstrate more favorable co-segregation in the compressive stress field near the dislocation core.

Figures \ref{fig7} e and f shows the binding energy $\Delta E_\text{bind}^{\text{GB,I-J}}$ distributions and the variation of $\Delta E_\text{bind}^{\text{GB,I-J}}$ in the dislocation core region, characterizing which solute pairs are more likely to maintain their pairing in this region and highlighting their stability and interaction tendencies at the LAGB. Ca-Ca pairs exhibit positive binding energies near the dislocation, with $\Delta E_\text{bind}^{\text{GB,I-J}}$ values decreasing, i.e. becoming less positive, as they approach closer to the core. Conversely, for Ca-Zn and Ca-Al solutes, the $\Delta E_\text{bind}^{\text{GB,I-J}}$ values  become less negative as they approach the core. Zn-Zn solutes display a behavior akin to Ca-Ca solutes, maintaining positive binding energy but with a gradual decrease near the dislocation core. 

\section{Discussion}
\begin{figure*}[ht!]
\centering
\includegraphics[width=\textwidth]{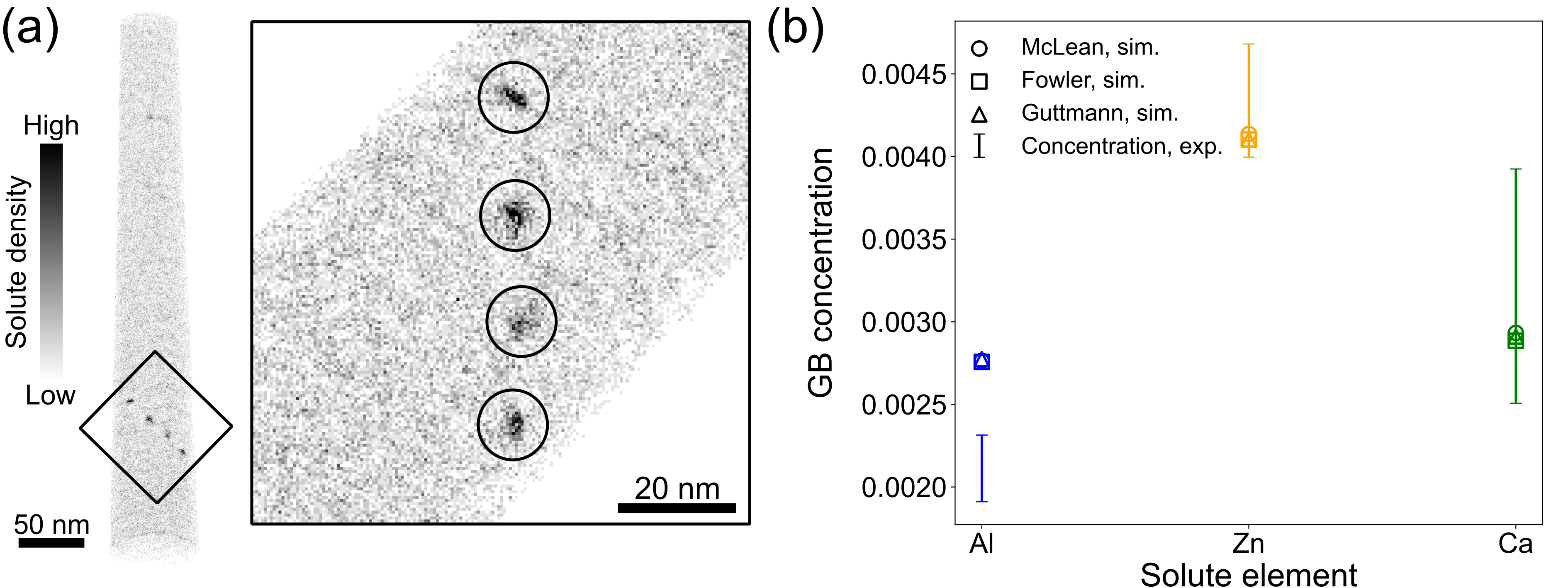}
\caption{Comparison of experimentally measured and theoretically predicted solute concentration at the LAGB. (a) In-plane solute distribution visualized along the dislocation array of the LAGB. (b) Solute concentration within 6 nm of the dislocation cores. Experimental bounds correspond to measurements from the four dislocation core regions marked in (a). Theoretical predictions of the GB solute concentrations are based on the McLean, Fowler, and Guttmann models, incorporating calculated segregation and solute-solute binding energies within 6 nm of the simulated dislocation core.}
\label{fig9}
\end{figure*}

In this study, APT-informed atomistic simulations were performed on a dilute AZX010 Mg alloy to explore the interplay between solute segregation and low-angle boundaries, aiming to unravel the underlying interaction mechanisms at the dislocation core level. APT played a crucial role in identifying compositionally distinct domains within the investigated LAGB and understanding their function in accommodating solute atoms. Through APT characterization, linear segregation patterns were revealed, indicating solute-decorated dislocation arrays, which occurred periodically along the boundary following the periodicity of the array of dislocations. Additionally, statistical nearest neighbor analysis of the reconstructed chemical and spatial information from the APT dataset highlighted quantifiable differences in solute clustering tendencies between the Mg matrix and the GB region. Among the explored segregation scenarios, mono-segregation of individual solutes to the LAGB is particularly relevant at low solute concentrations, where solute-solute interactions are negligible. The segregation characteristics of solute atoms to the dislocation array are determined by their binding with the dislocations, which is influenced by the misfit parameter. Al and Zn, which have a negative misfit relative to the host Mg atoms, showed favorable segregation on the compression side of the dislocation stress field. In contrast, Ca, with a positive misfit, segregates on the tension side. This trend is further supported by elastic dipole calculations, from which positive and higher-magnitude dipole values for Ca suggest a stronger interaction with tensile fields. The bi-modal distribution of segregation energies $\Delta E_\text{seg}$ at the LAGB aligns well with recent observations in basal-textured Mg polycrystals \cite{ganguly2024grain}, highlighting the role of tilt LAGBs as critical sites for inhomogeneous solute distribution.

The interactions between solutes at the LAGB reveal intriguing synergistic effects, particularly with Ca's attraction to Zn and Al. This is demonstrated by the increased occurrence of Ca-Zn and Ca-Al solute pairs, as well as more complex formations like Ca-Zn-Zn and Ca-Ca-Zn triplets, within the GB region of the APT reconstructed sample. The trend is driven not only by individual solutes preferentially clustering at the GB but also by their synergistic co-segregation into these specific pairings. Remarkably, both Ca-Ca pairs and Ca-Ca-Ca triplets show a pronounced clustering tendency at the GB compared to their distribution in the bulk, as highlighted by APT measurements. While the calculated binding energies in Figure \ref{fig7} indicate that Ca clusters are energetically unfavorable in the bulk material, corroborated by their sparse presence in Figure  \ref{fig5}d and e, their abundance increases dramatically at the GB, being approximately 2- (Ca-pairs) and 12-fold (Ca-triplets) higher than in the bulk. This significant enrichment can be attributed to their favorable clustering propensity at the GB, which is further supported by the energetic landscape. More negative $\Delta E_\text{seg}^{\text{I-J}\prime\prime}$ values suggest that solute pairs are energetically favored at the GB over the bulk, whereas less positive $\Delta E_\text{bind}^{\text{GB,I-J}}$ indicates a greater preference for solutes to remain clustered rather than disperse at the GB. As solute concentrations rise, it would be expected that synergistic co-segregation effects and clustering tendencies become more pronounced. However, the relatively low solute concentrations in the bulk, measured at 0.167 at.\% Al, 0.174 at.\% Ca and 0.333 at.\% Zn, mean that a mono-segregation scenario in accordance with Equation~\eqref{equ3} and Figures \ref{fig7}a and b adequately accounts for the observed segregation behavior. The $\Delta E_\text{seg}$ spectra provide a qualitative representation of the segregation ratios measured by APT, effectively capturing the dynamics of solute distribution in this context. 

Additionally, thermodynamic models for GB segregation were employed to quantitatively predict solute concentrations at the LAGB, using calculated segregation and pair interaction energies. These predictions were then compared to APT-measured values, as illustrated in Figure \ref{fig9}b. Experimentally determined solute concentrations within the dislocation core region (6 nm radius) were gathered across four distinct regions of interest, which are outlined in Figure \ref{fig9}a. Three models were employed: the McLean model, which assumes no solute-solute interactions (mono-segregation scenario); the Fowler model, which incorporates additional pair interaction contributions from identical solute elements; and the more comprehensive Guttmann model, which accounts for all pair interactions between different solute elements. The influence of pair interactions on predicted solute concentrations is minimal, as evidenced by the slight differences among models predictions, likely due to the overall low solute concentrations in this system. Overall, there is good agreement between predicted and experimental concentrations, with Al being an exception where its concentration is overestimated by the models.  

The broader implications of our findings underscore the significant impact of solute atom segregation at microstructural defects on material behavior and properties. The segregation of specific solute atoms at dislocations, such as Ca, Al, and Zn in this study, can induce delocalization of the core regions. This phenomenon occurs as the cores expand over a broader area, driven by structural and chemical reordering due to the presence of these atoms. The resultant lower energies are attributed to decreased localized lattice distortions, suggesting that the interaction energy between delocalized cores diminishes. This reduction in interaction energy offers several potential benefits for material properties. For instance, it could enhance slip characteristics by facilitating easier movement of dislocations through the material, thereby improving ductility and toughness. Additionally, reduced interaction energies may promote boundary migration, which is crucial in processes such as grain growth and recrystallization.

\section{Conclusions}

In summary, we utilized atomic-scale experimental methods and atomistic modeling to investigate the co-segregation mechanisms of Ca, Zn, and Al solutes at a low-angle grain boundary in a dilute AZX010 magnesium alloy. Three-dimensional atom probe tomography revealed pronounced enrichment of Ca, Zn, and Al at the LAGB, with Ca exhibiting a particularly strong clustering tendency along the linear dislocation array. Cluster analysis of the APT data indicated a significant increase in Ca-Ca pairs and Ca-containing clusters at the grain boundary compared to the matrix. Atomistic simulations showed that Ca solutes preferentially segregate to tensile regions around dislocation cores, while Zn and Al favor compressive regions, consistent with their respective atomic size misfits in Mg. Notably, the simulations demonstrated that the co-segregation of Ca-Ca pairs near the dislocation core is energetically more favorable than that of other solute pairs or individual solutes, providing an atomistic explanation for the experimental observations. Thermodynamic models incorporating calculated segregation and pair interaction energies predicted solute concentrations at the LAGB that agree well with experimental measurements, validating the modeling approach. 

The combined experimental and modeling findings underscore the critical importance of understanding synergistic solute interactions at the dislocation core level. This enhanced insight not only rationalizes the observed clustering behavior but also has broader implications for tailoring grain boundary properties and improving the mechanical performance of magnesium alloys through targeted alloying and grain boundary engineering strategies.

\section*{Acknowledgements}

R.P. and T.A.S. are grateful for the financial support from the German Research Foundation (DFG) (Grant Nr. AL1343/7-1, AL1343/8-1 and Yi 103/3-1). H.W., Z.X. and T.A.S. acknowledge the financial support by the DFG (Grant Nr. 505716422). Z.X. and S.K.K. acknowledge financial support by the DFG through the projects A05 and C02 of the SFB1394 Structural and Chemical Atomic Complexity – From Defect Phase Diagrams to Material Properties, project ID 409476157. Additionally, Z.X. and S.K.K. are grateful for funding from the European Research Council (ERC) under the European Union’s Horizon 2020 research and innovation programme (grant agreement No. 852096 FunBlocks). Fundings from the French National Research Agency (ANR) grant ANR-21-CE08-0001 (ATOUUM) and grant ANR22-CE92-0058-01 (SILA) are acknowledged by J.G., T.R., and by S.B., J.G., J.P., T.R., respectively. The authors gratefully acknowledge the computing time provided to them at the NHR Center NHR4CES at RWTH Aachen University (project number p0020267). This is funded by the Federal Ministry of Education and Research, and the state governments participating on the basis of the resolutions of the GWK for national high performance computing at universities (www.nhr-verein.de/unsere-partner).

\bibliographystyle{elsarticle-num}

\bibliography{main}

\end{document}